\documentclass[usenatbib,a4paper,times,fleqn]{mnras}
\usepackage{graphicx,natbib,times,amssymb,amsmath,pstricks,mathtools,enumitem,array}
\usepackage{aas_macros}
\usepackage{bm,url}
\usepackage{textcomp}
\usepackage{breakurl}
\usepackage{subfig}
\usepackage{nicefrac,xfrac}
\usepackage{mathtools}
\usepackage{cancel}
\usepackage{multicol}
\usepackage{listings}
\usepackage{cuted}

\newcommand {\apgt} {\ {\raise-.5ex\hbox{$\buildrel>\over\sim$}}\ }
\newcommand {\aplt} {\ {\raise-.5ex\hbox{$\buildrel<\over\sim$}}\ }

\title[\textsc{dustywave} at large dust fractions]{Large dust fractions can prevent the propagation of soundwaves}

\author[David--Cl\'eris \& Laibe]{Timoth\'ee David--Cl\'eris$^{1}$\thanks{timothee.david--cleris@ens-lyon.fr}, Guillaume Laibe$^{1,2}$\thanks{guillaume.laibe@ens-lyon.fr}- \\
$^{1}$Univ Lyon, Univ Lyon1, Ens de Lyon, CNRS, Centre de Recherche Astrophysique de Lyon UMR5574, F-69230, Saint-Genis,-Laval, France.\\
$^{2}$Institut Universitaire de France\\
}
\pagerange{\pageref{firstpage}--\pageref{lastpage}} \pubyear{2020}
\date{}
\begin{document}
\label{firstpage}
\bibliographystyle{mnras}
\maketitle

\begin{abstract}
Dust plays a central role in several astrophysical processes. Hence the need of dust/gas numerical solutions, and analytical problems to benchmark them. In the seminal dustywave problem, we discover a regime where sound waves can not propagate through the mixture above a large critical dust fraction. We characterise this regime analytically, making it of use for testing accuracy of numerical solvers at large dust fractions. 
\end{abstract}

\section{Introduction}
\label{sec:introduction}

Quantitative study of dust is of prime importance in astrophysics. Numerical simulations are used for determining the 3D evolution of dust/gas systems. Providing accurate tests to benchmark these numerical codes is therefore critical to ensure reliability of the results. Astrophysical dust is usually modelled by a pressureless continuum that exchanges momentum with the gas through a drag force \citep{Saffman1962,Baines1965,Clair1970,Marble1970}. Several analytical problems involving advection, waves, shocks, settling or dust/gas instabilities have been used to benchmark dust/gas codes (e.g. \citealt{BL2019,Stoyanovskaya2020} and references therein for recent discussions).

The \textsc{dustywave} problem consists of the propagation in 1D of a sound wave in such a mixture (\citealt{Ahuja1973,Gumerov1988,LP2011,dustywave2016} -- see Sect.~\ref{sec:dustywave}). \textsc{dustywave} is one of the most widely used benchmark, since it associates dust/gas drag and gas compressibility, both in Lagrangian  (e.g. \citealt{LP2012a,LP2014,LB2014,Booth2015,PL2015,Stoyanovskaya2018b,Mentiplay2020}) or in Eulerian methods (e.g. \citealt{Porth2014,Yang2016,Hubber2018,McKinnon2018,Riols2018,Lebreuilly2019,Moseley2019}). In attempting to benchmark a numerical code against this test, we figured out the existence of a regime at large dust-to-gas ratios where waves cannot propagate. After having recalled briefly the main properties the \textsc{dustywave} problem in Sect.~\ref{sec:dustywave}, we show the existence of this regime, derive analytic values for the corresponding boundaries, provide physical explanations and numerical tests in Sect.~\ref{sec:bifurc}. 

\section{Dispersion relation}
\label{sec:dustywave}

The equations of evolution for a 1D astrophysical dusty mixture are
	\begin{align}
	\partial_{t} \rho_{\rm g} + {v}_{\rm g}  \partial_x \rho_{\rm g} &=  - \rho_{\rm g} \,  \partial_x  {v}_{\rm g} \label{eq:syst1}, \\
		\partial_{t} \rho_{\rm d} + {v}_{\rm d}  \partial_x \rho_{\rm d} &=  - \rho_{\rm d} \,  \partial_x  {v}_{\rm d} , \\
		\partial_{t} {v}_{\rm g} + {v}_{\rm g}  \partial_x {v}_{\rm g} &=  + \frac{K}{\rho_{\rm g}} \left( {v}_{\rm d} - {v}_{\rm g} \right) - \frac{\partial_x P}{\rho_{\rm g}} , \\
		\partial_{t} {v}_{\rm d} + {v}_{\rm d}  \partial_x {v}_{\rm d}  &=  - \frac{K}{\rho_{\rm d}} \left( {v}_{\rm d} - {v}_{\rm g} \right) , \label{eq:syst4}
	\end{align}
where $\rm g$ and $\rm d$ stand for gas and dust respectively (e.g. \citealt{Garaud2004}) and $K$ denotes the drag coefficient. Assuming isothermal gas $ P = c_{\rm s}^2 \rho_{\rm g} $, we expand linearly Eqs.~\ref{eq:syst1} -- \ref{eq:syst4} under the generic form $a = a_{0} + \delta a$, with $v_{\mathrm{d}0} = v_{\mathrm{g}0} = 0 $. One obtains
	\begin{align}
		\partial_{t} \delta \rho_{\rm g} &= - \rho_{\mathrm{g}, 0} \,  \partial_x \delta v_{\rm g} , \label{eq:systlin1} \\
		\partial_{t} \delta \rho_{\rm d} &= - \rho_{\mathrm{d}, 0}  \, \partial_x \delta v_{\rm d}, \\
		\partial_{t} \delta v_{\rm g} &= + \frac{K}{\rho_{\mathrm{g}, 0}} \left( \delta v_{\rm d} - \delta v_{\rm g} \right) - c_{\rm s}^2 \frac{\partial_x \delta \rho_{\rm g}}{\rho_{\mathrm{g}, 0}} , \\
		\partial_{t} \delta v_{\rm d} &= - \frac{K}{\rho_{\mathrm{d}, 0}} \left( \delta v_{\rm d} - \delta v_{\rm g} \right)  . \label{eq:systlin4}
	\end{align}
We decompose the perturbation on Fourier space under the form $\delta a = \tilde{a} e^{i(k x - \omega t)}$ for each perturbed field, giving the condition
\begin{equation}
\begin{vmatrix}
 -i \omega & 0 & \frac{i c_{\rm s} k \rho_{\mathrm{g}, 0}}{\rho_{\mathrm{g}, 0} + \rho_{\mathrm{d}, 0} } & 0 \\
 0 & -i \omega & 0 & \frac{i c_{\rm s} k \rho_{\mathrm{d}, 0}}{\rho_{\mathrm{g}, 0} + \rho_{\mathrm{d}, 0}  } \\
 \frac{i c_{\rm s} k \left( \rho_{\mathrm{g}, 0} + \rho_{\mathrm{d}, 0} \right) }{\rho_{\mathrm{g}, 0}} & 0 &-i \omega + \frac{1}{t_\mathrm{g}} & -\frac{1}{t_\mathrm{g}} \\
 0 & 0 & -\frac{1}{t_\mathrm{d}} &-i \omega + \frac{1}{t_\mathrm{d}} 
\end{vmatrix}
 = 0 .
\end{equation}

    \begin{figure}
    \includegraphics[width=\columnwidth]{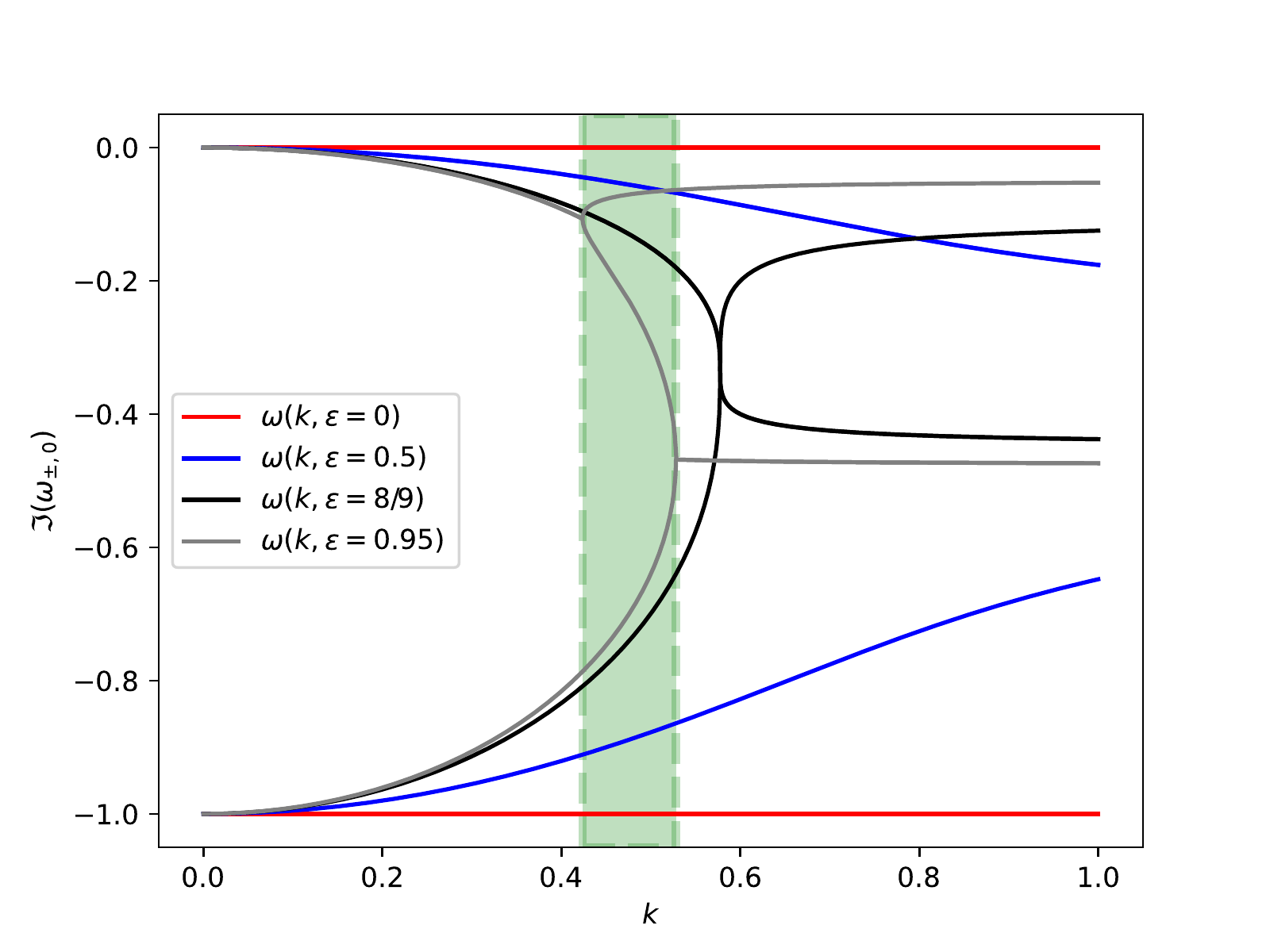}
    \caption{Imaginary parts of the roots $\omega_{\pm}$ and $\omega_{0}$ for different values of the dust fraction $\epsilon$ revealing a bifurcation at $\epsilon \ge \epsilon_{\rm c} = 8/9$.}
    \label{fig:im_part} 
    \end{figure}

We obtain the following dispersion relation
\begin{equation}
\omega^4  +\frac{i}{ t_{\rm s}} \omega^3-c_{\rm s}^2 k^2 \omega^2-\frac{i}{ t_{\rm s}} c_{\rm s}^2 k^2  (1-\epsilon ) \omega = 0 ,
\label{eq:disp_dim}
\end{equation}

where the barycentric stopping time is $t_{\rm s} \equiv  \frac{\rho_{\mathrm{g}, 0} \rho_{\mathrm{d}, 0}}{K \left( \rho_{\mathrm{g}, 0} + \rho_{\mathrm{d}, 0}\right)}$ and $\epsilon \equiv \frac{\rho_{\mathrm{d}, 0}}{\rho_{\mathrm{g}, 0} + \rho_{\mathrm{d}, 0}}$ is the total dust fraction. Rescaling time and space by $t_{\rm s}$ and $c_{\rm s} t_{\rm s}$ respectively gives in the dimensionless form
\begin{equation}
\omega^4  +i \omega^3- k^2 \omega^2- i k^2  (1-\epsilon )\omega = 0 ,
\label{eq:disp_full}
\end{equation}
where we preserved the notations $\omega$ and $k$ for further readability. We disregard the solution $\omega_{\rm null} = 0$ on the null space. On the column space, Eq.~\ref{eq:disp_full} reduces to
\begin{equation}
\omega^3  +i \omega^2- \omega k^2- i k^2  (1-\epsilon ) = 0 ,
\label{eq:disp}
\end{equation}
which can alternatively be written under the convenient form.
\begin{equation}
\omega^2- k^2  +\frac{i}{\omega} \left(\omega ^2- k^2  (1-\epsilon )\right)= 0 .
\label{eq:disp_dim}
\end{equation}
The change of variable $\omega = i y$ gives a cubic with real positive coefficients
\begin{equation}
y^3  +y^2 + y k^2 + k^2  (1-\epsilon ) = 0 .
\label{eq:disp_real}
\end{equation}
When Eq.~\ref{eq:disp_real} admits two complex conjugated roots and one real root, the latter is negative since $k^2  (1-\epsilon ) > 0$. When Eq.~\ref{eq:disp_real} admits three real roots, Descarte's rule of sign shows that they are all negative. Since $\Im \left( \omega \right) = \Re\left( y\right)$, all modes of the \textsc{dustywave} problem are damped. This result can alternatively be found from the argument principle \citep{Debras2020}. Let split $\omega$ in its real and imaginary part by setting $\omega \equiv r + i s$. One obtains
    \begin{align}
        0&=r \left[ r^2 - (3 s^2 +2 s+ k^2)\right] , \label{eq:syst1a}\\
        0&=s^3 + s^2 + s (k^2 - 3 r^2) + (k^2 (1-\epsilon) - r^2) . \label{eq:syst1b}
    \end{align}
Eq.~\ref{eq:syst1a} shows that the three expected modes decompose as follow: 
    \begin{align}
        r&=0 , \label{eq:syst2a}\\
        0&=s^3 + s^2 + s k^2  +k^2 (1-\epsilon) , \label{eq:syst2b}
    \end{align}
and
    \begin{align}
        r^2 &=3 s^2 +2 s+ k^2 , \label{eq:syst3a}\\
        0&=s^3 + s^2 + s \frac{1}{4}\left( k^2 +1\right)  +k^2 \frac{\epsilon}{8} . \label{eq:syst3b}
    \end{align}
We note that the symmetry $r \to -r$ implies three solutions $\omega_{\pm} = \pm \vert r \vert +i s$ and $\omega_{0} = i s$.  Eqs.~\ref{eq:syst2a} -- \ref{eq:syst2b} give solutions that are always purely damped. Eqs.~\ref{eq:syst3a} -- \ref{eq:syst3b} give solutions that are contra-propagative and damped.

\section{Absence of sound propagation}
\label{sec:bifurc}

\subsection{Analysis}

    \begin{figure}
    \includegraphics[width=\columnwidth]{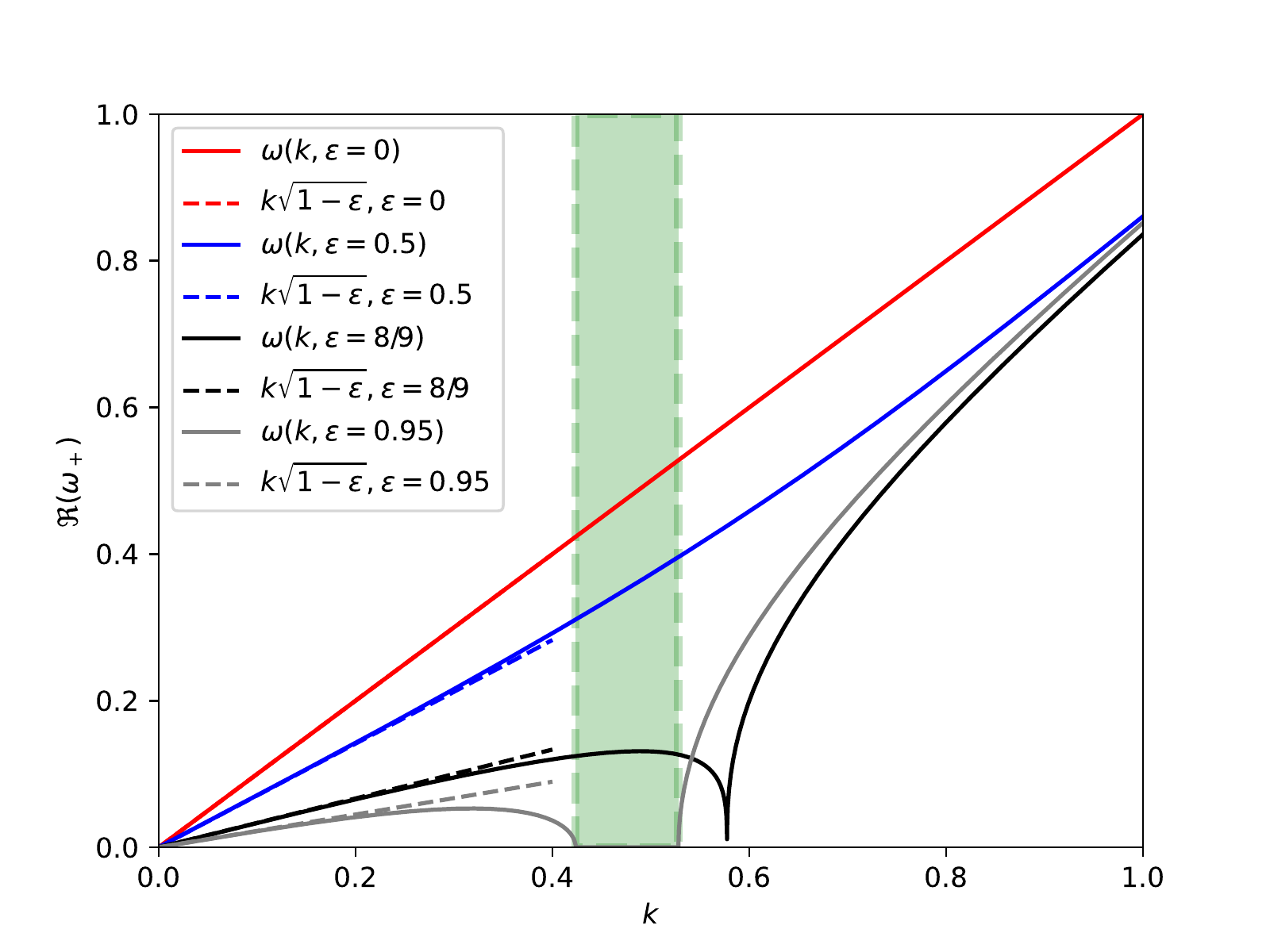}
    \caption{Real part of the root $\omega_+$ for different values of the dust fraction $\epsilon$ (solid lines) superimposed to the asymptotic regimes (dashed lines). A forbidden band develops for $\epsilon \ge \epsilon_{\rm c} = 8/9$.}
    \label{fig:real_part} 
    \end{figure}

\begin{figure*}
    \includegraphics[width=\textwidth]{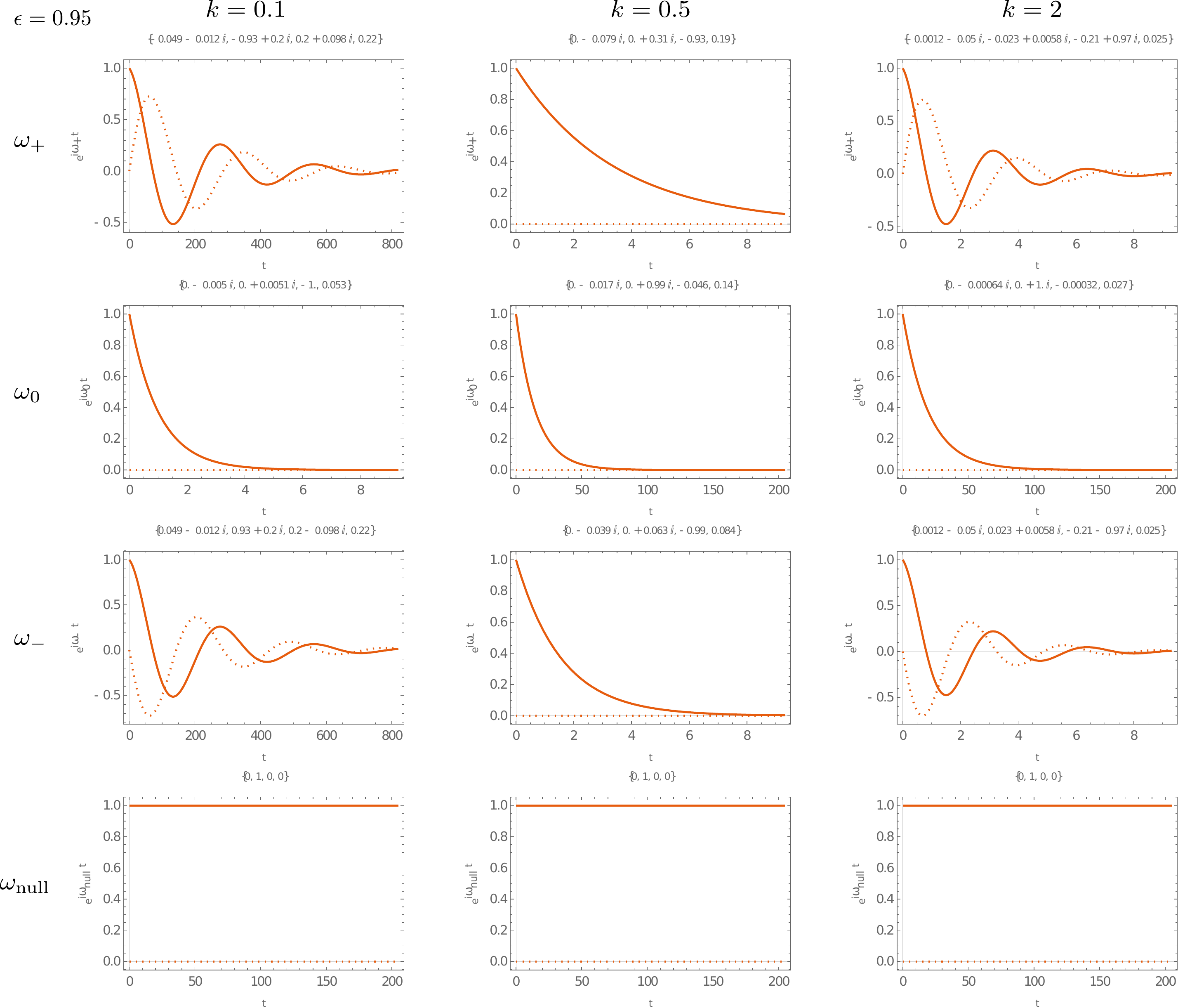}
    \caption{Real (solid line) and imaginary (dotted line) parts of $\mathrm{e}^{i \omega t}$ for the modes $\omega_{+}$, $\omega_{0}$ and $\omega_{-}$ ($\epsilon  = 0.95$). The middle case $k = 0.5$ corresponds to the regime where no mode propagates.  Note the different timescales used for the plot for readability. The coefficients of the eigenvectors corresponding to the eigenvalues are provided on top of the plots and can be found in Table.~\ref{table:eigenvalues}--\ref{table:eigenvectors}. The mode $\omega_{\rm null}$ corresponding to the null space of Eqs.~\ref{eq:systlin1} -- \ref{eq:systlin4} is provided for sanity check.}
    \label{fig:evol_high_eps} 
    \end{figure*}

Fig.~\ref{fig:real_part} shows that sound waves behave in a specific manner at large dust fractions. Above a critical value $\epsilon_{\rm c}$, an interval of values for $k$ where waves can not propagate develops (a so-called forbidden band). We relate the existence of the bifurcation at $\epsilon_{\rm c}$ to the fact that above this value, all three solutions come solely from Eq.~\ref{eq:syst2b}.   From the discriminant of the cubic Eq.~\ref{eq:disp}, one gets the condition $\epsilon \geq \epsilon_{\rm c} = 8/9$ and $k_{\rm min} 
\leq k \leq k_{\rm max}$ with
\begin{equation}
   \begin{cases}
      k_{\min} & \equiv \displaystyle \sqrt{\left(1-\frac{9 \epsilon }{8}\right) \left(3 \epsilon  -\frac{4}{3} +  \sqrt{\epsilon }  \sqrt{9 \epsilon -8}\right)+\frac{1}{3}} ,\\
  k_{\max} &  \equiv \displaystyle \sqrt{\left(1-\frac{9 \epsilon }{8}\right) \left(3 \epsilon  -\frac{4}{3} -  \sqrt{\epsilon }  \sqrt{9 \epsilon -8}\right)+\frac{1}{3}} ,\\
   \end{cases}
   \label{eq:gas_mode}
  \end{equation}
The centre of the band $k_{\rm c}$ can be estimated from the relation 
\begin{equation}
k_{\rm c}^{2} \equiv \sqrt{k^{2}_{\min}k^{2}_{\max}} = \sqrt{1 - \epsilon},
\end{equation}
which indicates that the band is centred around $k_{\rm c} \sim \left( 1 - \epsilon_{\rm c} \right)^{1/4} = 1/\sqrt{3} \sim 0.57 $, except for values of $\epsilon$ extremely close to unity (see Sect.~\ref{sec:physics} for the physical explanation). In that regime, this condition gives $r^2 < 0$ in Eq.\ref{eq:syst2b}. This ensures three and only three complex roots for the dispersion relation as expected. Another quick way to find the critical value $\epsilon_{\rm c} = 8/9$ consists of solving for $k$ as function of $s$ in Eq.~\ref{eq:syst2b} to get $k(s)^2 = - \frac{2s + 8s^2 + 8s^3}{2s + \epsilon}$, which gives by enforcing $r^{2} < 0$ in Eq.~\ref{eq:syst2b}, $-2 s^2 + (3 \epsilon -4) s - (1-\epsilon) > 0$. Positivity is ensured for positive discriminant, i.e. 
    \begin{equation}
    \Delta = \epsilon (-8+9\epsilon) \geq 0 \Rightarrow \epsilon \ge \frac{8}{9} .
    \end{equation}
At the critical value $\epsilon = \epsilon_{\rm c}$, $k = 1/\sqrt{3}$ and the dispersion relation Eq.~\ref{eq:disp} factorises according to $\left( \omega + i /3 \right)^{3} = 0$. Real and imaginary parts of $\mathrm{e}^{i \omega t}$ for the different modes at $\epsilon = 0.95 > \epsilon_{\rm c}$ are shown on Fig.~\ref{fig:evol_high_eps}. As expected, at $k = 0.5$, no mode propagates. This plot can be compared to the case $\epsilon = 0.1$, where the modes $\omega_{\pm}$ propagate for any value of $k$ (Fig.~\ref{fig:evol_low_eps}, Appendix~\ref{app:low_eps}).

\subsection{Eigenvectors}

\begin{figure*}
    \includegraphics[width=\textwidth]{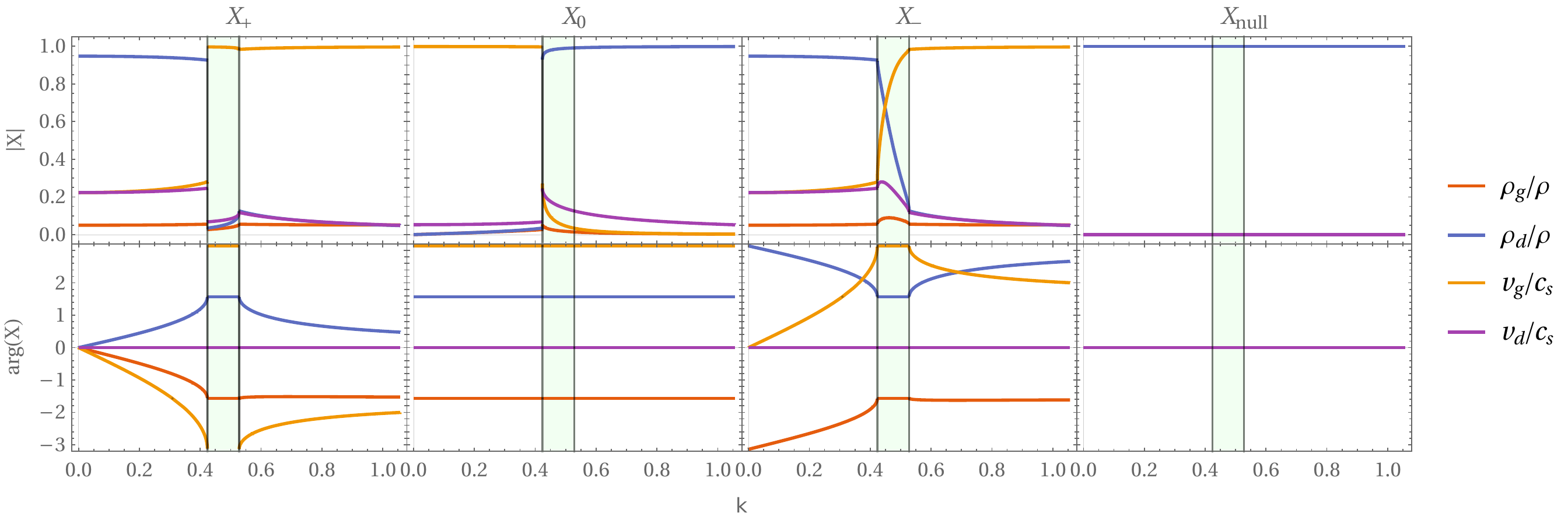}
    \caption{Modulus (top) and phase (bottom) of the eigenvectors at the \textbf{bifurcation} for the modes $\omega_{+}$, $\omega_{0}$ and $\omega_{-}$ and $\omega_{\rm null}$ (from left to right, $\epsilon  = 0.95$). Values of the related eigen system can be found in Tables.~\ref{table:eigenvalues} -- \ref{table:eigenvectors}. }
    \label{fig:eigenmodes} 
    \end{figure*}

Fig.~\ref{fig:eigenmodes} shows the modulus (top) and the argument (bottom) of the eigenvectors corresponding to the modes $\omega_{+}$, $\omega_{0}$, $\omega_{-}$ and $\omega_{\rm null}$ for values of $k$ centred around the bifurcation. The amplitudes of the modes $\omega_{\rm +}$ and $\omega_{-}$ are similar inside the forbidden band, although they are distinct outside. Differential phases of the mode $\omega_{0}$ do not depend on $k$, outside or inside the band. Differential phases of the modes $\omega_{\pm}$ are also constant inside the band. This ensures the required differential velocity from which the modes are damped. The mode $\omega_{\rm null}$ corresponds to a steady static perturbation on the dust density only, i.e. $\delta \rho_{\rm g} = 0$, $ \delta  v_{\rm g} = \delta  v_{\rm d} = 0 $.

\subsection{Physical interpretation}
\label{sec:physics}

In the limit $\epsilon \to 0$, the dispersion relation Eq.~\ref{eq:disp} reduces to
\begin{equation}
\omega^{2} + i\omega - k^{2} = 0 .
\label{eq:inter}
\end{equation} 
Eq.~\ref{eq:inter} admits the solutions $\tilde{\omega}_{\rm} = -i/2 \pm \sqrt{k^{2} - 1/4}$, showing the existence of $k_{\max} =1/2$ below which waved are damped (in this limit, $k_{\rm min} = 0$). Physically, Eq.~\ref{eq:inter} is also the dispersion relation associated to the reduced system
	\begin{align}
		\partial_{t} \delta \rho_{\rm g} &= - \rho_{\mathrm{g}, 0} \partial_{x} \delta v_{\rm g} , \label{eq:syst_simp1} \\
		\partial_{t}\delta v_{\rm g} &= - \frac{K}{\rho_{\mathrm{g}, 0}} \delta v_{\rm g} - c_{\rm s}^2 \frac{\partial_{x} \delta \rho_{\rm g}}{\rho_{\mathrm{g}, 0}} ,\label{eq:syst_simp2}
	\end{align}
where the gas dissipates its energy in a passive dust phase through back-reaction. When $\left(1 - \epsilon\right)$ is small but finite, large values of $k^{2}/\omega$ can satisfy the dispersion relation, which reduces to $i \omega^2- \omega k^2- i k^2  (1-\epsilon ) = 0$. This corresponds to propagation through the mixture and explains the finite value of $k_{\rm min}$  (e.g. \citealt{Laibe2014} and App.~\ref{sec:asympt}).

For the \textsc{dustywave} problem with multiple dust species, the regime identified in this study exists when the stopping times of the different species are close enough (we verified this fact  numerically). When this condition is not fulfilled, gas/mixture mode can propagate through one specie even if it can not through the other one, and no forbidden band is expected. This example provides a further situation where a physical effect observed in a mixture with a single grain size does not occur when several sizes are considered (e.g. \citealt{Krapp2019}).

\section{conclusion}
\label{sec:conclusion}

We identified and characterised the existence of a regime in the single specie \textsc{dustywave} problem where waves can not propagate, neither as a gas mode or a mixture mode. This regime develops above $\epsilon_{\rm c} = 8/9$, when the dust-to-gas ratio is sufficiently large for gas to dissipate its energy in an independent dust phase via back-reaction. Numerical solvers can be checked in this regime to verify their accuracy at large dust fractions.

\section*{Acknowledgements}

GL acknowledges funding from the ERC CoG project PODCAST No 864965. This project has received funding from the European Union's Horizon 2020 research and innovation programme under the Marie Sk\l odowska-Curie grant agreement No 823823. This project was partly supported by the IDEXLyon project (contract nANR-16-IDEX-0005) under the auspices University of Lyon. We acknowledge financial support from the national programs (PNP, PNPS, PCMI) of CNRS/INSU, CEA, and CNES, France. We used \textsc{Mathematica} \citep{Mathematica}. We thank the anonymous referee for a thorough and insightful report and for suggesting the title.

\section*{Data availability}
All relevant data are given in the article.

\begin{appendix}
 \section{Asymptotic behaviours}
\label{sec:asympt}

When $k \gg 1 $ (the physical wavelength $\lambda \equiv 2 \pi c_{\rm s} t_{\rm s}/ k$ is smaller than the physical stopping length  $ c_{\rm s} t_{\rm s}$), Eqs.~\ref{eq:syst2a} -- \ref{eq:syst3b} provide

\begin{equation}
    \begin{cases}
        \omega_+ & = \left[ + \sqrt{k^{2} + 3\epsilon^{2}/ 4 - \epsilon} + \mathcal{O}\left( k^{-2} \right)  \right] - i \left[  \epsilon/2 \hspace{13.7pt} + \mathcal{O}\left( k^{-2} \right) \right] ,\\
        \omega_0 & = \hphantom{ \left[ + \sqrt{k^{2} + 3\epsilon^{2}/ 4 - \epsilon} + \mathcal{O}\left( k^{-2} \right)  \right] } - i  \left[  (1-\epsilon) + \mathcal{O}\left( k^{-2} \right) \right] , \\
        \omega_- & =  \left[ - \sqrt{k^{2} + 3\epsilon^{2}/ 4 - \epsilon} + \mathcal{O}\left( k^{-2} \right)   \right]-i\left[  \epsilon/2 \hspace{13.7pt} + \mathcal{O}\left( k^{-2} \right) \right] ,
    \end{cases}
\end{equation}
or in a even more simplified form,
\begin{equation}
    \begin{cases}
        \omega_+ & = \left[ + k  + \mathcal{O}\left( 1 \right) \right] - i \left[  \epsilon/2 \hspace{13.7pt} + \mathcal{O}\left( k^{-2} \right) \right] ,\\
        \omega_0 & = \hphantom{\left[ + k  + \mathcal{O}\left( 1 \right) \right]  } - i  \left[  (1-\epsilon) + \mathcal{O}\left( k^{-2} \right) \right] , \\
        \omega_- & =  \left[ - k  + \mathcal{O}\left( 1 \right) \right]-i\left[  \epsilon/2 \hspace{13.7pt} + \mathcal{O}\left( k^{-2} \right) \right] .
    \end{cases}
\end{equation}
The evolution of the plane wave is correctly described by expanding both the real and the imaginary parts of $\omega$ to their respective leading orders. In this gas regime, the propagation is supported by the pressure of the gas and damped by dust back-reaction. Since $\Re \left(w_{\pm} \right)  = \pm k  + \mathcal{O}\left( 1 \right)$, the typical physical oscillation time is $\sim \lambda / c_{\rm s} \ll t_{\rm s}$. After a typical time $\sim (1 - \epsilon)^{-1}t_{\rm s}$, the initial dust velocity adjusts onto the one of the gas (mode $\omega_{0}$). Meanwhile, the gas undergoes several oscillations that are supported by its own pressure (terms $\pm k$, modes $\omega_{\pm}$) that are progressively damped by dust back-reaction (terms $-i\epsilon/2$, the factor 2 accounting for dissipation by both modes). In the limit $k \ll 1 $, corresponding to $\lambda \gg c_{\rm s} t_{\rm s}$, one obtains
\begin{equation}
    \begin{cases}
        \omega_+ &= + \sqrt{1-\epsilon} k  - i \displaystyle \epsilon k^2 / 2   \hspace{21.1pt} + \mathcal{O} \left(k^{3} \right) , \\
        \omega_0 &= \hphantom{+ \sqrt{1-\epsilon} k} +i (-1 + \epsilon k^2) + \mathcal{O} \left(k^{3} \right) ,\\
        \omega_- &= - \sqrt{1-\epsilon} k - i \displaystyle \epsilon k^2 /2  \hspace{21.1pt} + \mathcal{O} \left(k^{3} \right) .
    \end{cases}
  \label{eq:mixture_mode}
\end{equation}
In this mixture regime, the propagation is supported by both gas and dust simultaneously, and damped by an effective diffusion. The typical oscillation time for a perturbation satisfies $\lambda / c_{\rm s} \gg t_{\rm s}$. After a typical time $t_{\rm s}$, the gas and dust  velocities have relaxed towards the barycentric velocity of the mixture (mode $\omega_{0}$, factor $-i$). Since the stopping time is much shorter than the oscillation time, the drag maintains the two phases well-coupled and the system tends to oscillate at the sound speed of the mixture $c_{\rm s} \sqrt{1 - \epsilon}$, which accounts for the inertia of the dust \citep{LP2012a}. Damping comes from the $ - i \displaystyle \epsilon k^2 /2$ term, which originates from the effective diffusion of the terminal velocity approximation \citep{Laibe2014}.

\section{Parameters for numerical tests}

We provide parameters for numerical test -- before ($k = 0.1$), at ($k = 0.5$) and after ($k = 2.$) the bifurcation at $\epsilon = 0.95$. The eigenvalues $\omega_{+}$, $\omega_{0}$ and $\omega_{-}$ are given in Table.~\ref{table:eigenvalues}. The corresponding values for the eigenvectors are given in Table.~\ref{table:eigenvectors}  Evolution of the densities and velocities are shown on Fig.~\ref{fig:evol_high_eps}.

\begin{table*}
\begin{center}
\begin{tabular}{|l|c|c|c|}
  \hline
$\omega$  & $k = 0.1$  & $k = 0.5$ & $k = 2.$  \\
  \hline
$\omega_{+}$ & $0.0219514\, -0.00479354 i$ &  $-0.639778 i$ & $1.93044\, -0.474696
   i$ \\
$\omega_{0}$  &  $-0.990413 i$ &  $-0.0665243 i$ &  $-0.0506079 i$\\
 $\omega_{-}$ &  $-0.0219514\, -0.00479354 i$ &  $-0.293697 i$ & $-1.93044\, -0.474696 i$\\
 \hline
  \end{tabular}
\caption{Eigenvalues for numerical tests of the \textsc{dustywave} regime ($\epsilon = 0.95$).}
\label{table:eigenvalues}
\end{center}
\end{table*}

\begin{table*}
\begin{center}
\begin{tabular}{|c|c|c|c|}
  \hline 
$^\top {X} = \left(\frac{\delta \rho_{\rm g}}{\rho_{0}} , \frac{\delta \rho_{\rm d}}{\rho_{0}} , \frac{\delta v_{\rm g}}{c_{\rm s}} , \frac{\delta v_{\rm d}}{c_{\rm s}} \right) $ & $k = 0.1$  & $k = 0.5$ & $k = 2.$  \\
  \hline
$X_{+}$ 		
    & $\left( \begin{array}{c}0.0117633\, +0.0486907 i \\ -0.20202+0.925123 i\\0.098324\, +0.202488 i\\ 0.223959 i\end{array} \right) $ 
    & $\left( \begin{array}{c}0.0388307\\-0.0625475\\ -0.993722 i\\ 0.0842453 i\end{array} \right) $ 
    & $  \left( \begin{array}{c}0.0501937\, +0.00123333 i\\-0.00576226+0.0234334 i\\0.974816\, -0.214459 i\\ 0.0252485 i\end{array} \right)$
    \\
\hline$X_{0}$  		& $ \left( \begin{array}{c}0.00504115\\-0.00509254\\ -0.998564 i\\ +0.0530918 i\end{array} \right)$ & $\left( \begin{array}{c}0.0172053\\-0.98915\\ -0.0457829 i\\ +0.138532 i\end{array} \right) $& $\left( \begin{array}{c}0.000639653\\-0.999645\\ -0.000323715 i\\ +0.0266263 i\end{array} \right)$\\
\hline$X_{-}$  		& $\left( \begin{array}{c}0.0117633\, -0.0486907 i\\-0.20202-0.925123 i\\-0.098324+0.202488 i\\ +0.223959 i\end{array} \right) $& $\left( \begin{array}{c}0.0790522\\-0.308168\\ -0.928696 i\\+0.190543 i\end{array} \right) $&
  $ \left( \begin{array}{c}0.0501937\, -0.00123333 i\\-0.00576226-0.0234334 i\\-0.974816-0.214459 i\\ +0.0252485 i\end{array} \right) $\\
\hline $X_{\rm null}$  	& $ \left( \begin{array}{c}0\\i\\0\\0\end{array} \right)$ & $\left( \begin{array}{c}0\\i\\0\\0\end{array} \right)$ &$ \left( \begin{array}{c}0\\i\\0\\0\end{array} \right) $
  \\\hline
  \end{tabular}
\caption{Eigenvectors for numerical tests of the \textsc{dustywave} regime ($\epsilon = 0.95$).}
\label{table:eigenvectors}
\end{center}
\end{table*}

\section{Evolution at low $\epsilon$}
\label{app:low_eps}

Evolution of the perturbations are given for $\epsilon = 0.10 < \epsilon_{\rm c}$, for a purpose of comparison with Fig.~\ref{fig:evol_high_eps}. The regime where no mode propagates is not observed as expected.

    \begin{figure*}
    \includegraphics[width=\textwidth]{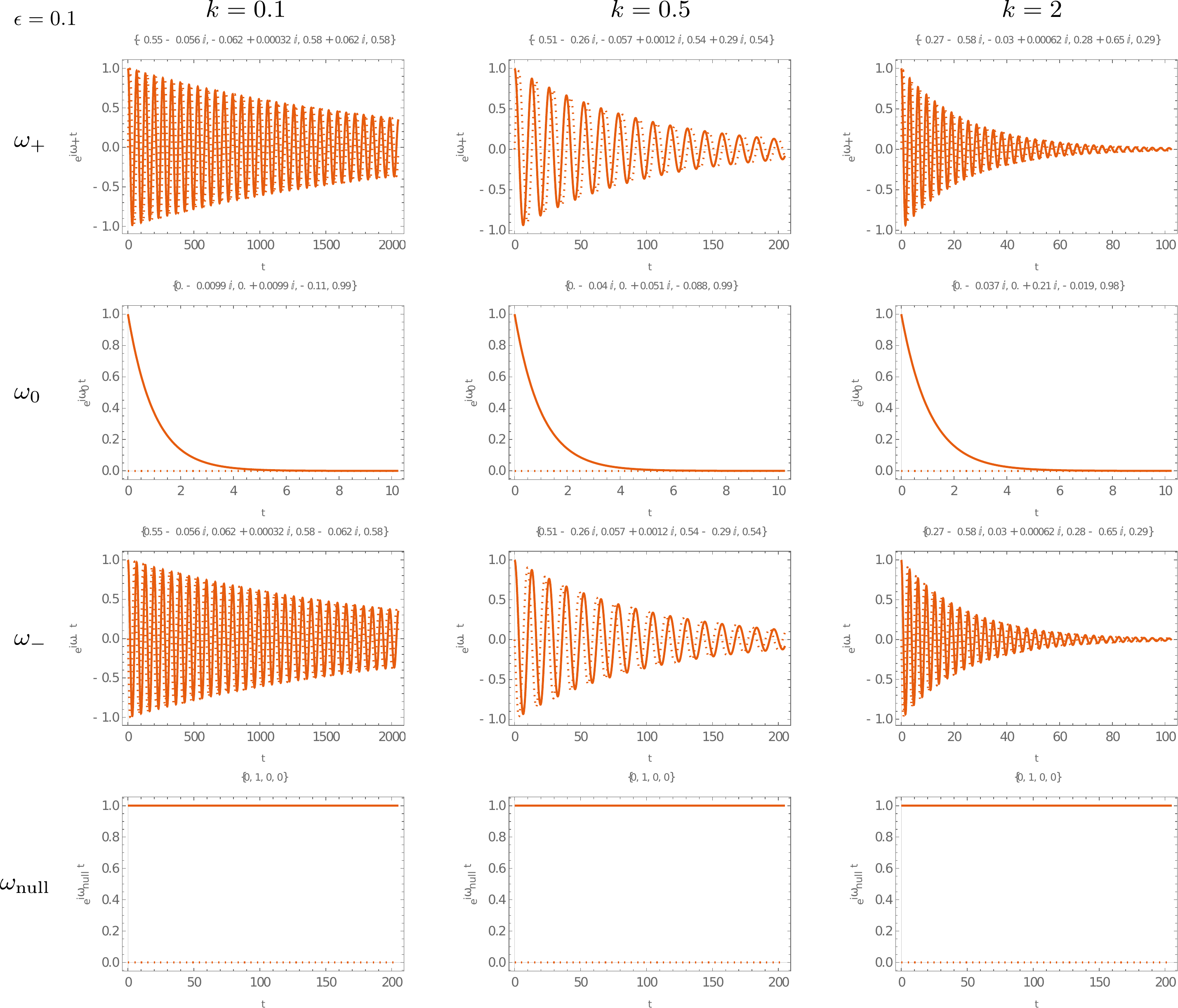}
    \caption{Real (solid line) and imaginary (dotted line) parts of $\mathrm{e}^{i \omega t}$ for the modes $\omega_{+}$, $\omega_{0}$ and $\omega_{-}$ and $\omega_{\rm null}$ ($\epsilon  = 0.10$). No regime is observed in this case and at least 2 modes propagate through the mixture.}
    \label{fig:evol_low_eps} 
    \end{figure*}

\end{appendix}

\label{lastpage}
\bibliography{dustywave}

\end{document}